\documentclass[11pt]{article}
\usepackage{setspace}
\usepackage{amsmath}
\usepackage{amssymb}
\usepackage{latexsym}
\usepackage{parskip}
\usepackage{cases}
\usepackage{mathrsfs} 
\usepackage[a4paper,left=1in,right=1in,top=1in,bottom=1in]{geometry}
\usepackage{graphicx} 
\usepackage{subfigure}
\usepackage{float}
\usepackage{enumerate}
\usepackage{listings}
\usepackage{color}
\usepackage{booktabs} 
\usepackage{setspace}
\usepackage{fancyhdr}
\author{}
\title{{\Large \textbf{Enhanced Monte Carlo Estimation of the Fisher Information Matrix with Independent Perturbations for Complex Problems}}}
\begin{document}
    \begin{center}
	\vspace*{1cm}
	
	\Large
	\textbf{Enhanced Monte Carlo Estimation of the Fisher Information Matrix with Independent Perturbations for Complex Problems}
	
	\vspace{0.5cm}
	\normalsize
	\textbf{Xuan Wu}
	\vspace{0.5cm}

	\textbf{Department of Applied Math and Statistics\\
	The Johns Hopkins University\\
	Baltimore, Maryland 21218, USA\\}
\vspace{0.5cm}
\end{center}
\large
\textbf{Abstract}

\quad The Fisher information matrix provides a way to measure the amount of information given observed data based on parameters of interest. Many applications of the FIM exist in statistical modeling, system identification, and parameter estimation. We sometimes use the Monte Carlo-based method to estimate the FIM because its analytical form is often impossible or difficult to be computed in real-world models. In this paper, we review the basic method based on simultaneous perturbations and present an enhanced resampling-based method with independent simultaneous perturbations to estimate the Fisher information matrix. We conduct theoretical and numerical analysis to show its accuracy via variance reduction from $O(1/N)$ to $O(1/(nN))$, where $n$ is the sample size of the data and $N$ is a measure of the Monte Carlo averaging. We also consider the trade-off between accuracy and computational cost. 
\section{Introduction}

\quad The Fisher Information Matrix (FIM) summarizes the amount of information about data related to parameters of interest, bearing in mind that the central role of data analysis is to extract information from data. We can see that the FIM plays a critical role in the theory and practice of statistical modeling, including system identification \cite{b1}, parameter estimation \cite{b2}, information theory \cite{b3}, and other areas. Let $\pmb{Z}$ be the collection of $n$ random vectors $[\pmb{z}_1,\ldots,\pmb{z}_n]$. In the following discussion, assume that $p(\pmb{Z}|\pmb{\theta})$ is the density function for the model of interest, where $\pmb{\theta}$ represents the unknown $p\times 1$ parameter vector, and $L(\pmb{\theta}|\pmb{Z})$ is the log-likelihood function of $p(\pmb{Z}|\pmb{\theta})$. Then the $p\times p$ FIM is defined as: 
$$\pmb{F}(\pmb{\theta})=\mathbb{E}\bigg[\frac{\partial L}{\partial \pmb{\theta}}\frac{\partial L}{\partial \pmb{\theta}^T}\bigg].$$ 

In general, it is difficult to compute the expectation of a product of nonlinear score functions with multiple parameters. On the other hand, if the second partial derivatives of $L$ exists, i.e., the Hessian matrix $\pmb{H}(\pmb{Z}|\pmb{\theta})$, then under specific regularity conditions \cite{b4}, the FIM can be represented as:
$$\pmb{F}(\pmb{\theta})=-\mathbb{E}\bigg[\frac{\partial^2 L}{\partial \pmb{\theta}\partial \pmb{\theta}^T}\bigg]=-\mathbb{E}[\pmb{H}(\pmb{\theta})],$$ 
which is generally easier to calculate compared with the basic definitional form.

In that there are no closed forms for $\pmb{F}(\pmb{\theta})$ in many practical problems and given the importance of the FIM, one way is to use averages of the estimated Hessian matrix based on pseudo data as the Monte Carlo estimation of the FIM. Moreover, some research studies present methods to estimate the FIM in general or specific cases (e.g., \cite{b5}, \cite{b6}). In particular, we can estimate the FIM via measurements of a Hessian matrix based on the simultaneous perturbation approach. Spall in \cite{b5} discusses a basic resampling-based approach. Later several papers present improved approaches, including a feedback-based approach in \cite{b7} and a method where information on some of the FIM is available in \cite{b8}.  In this paper, we introduce another enhanced Monte Carlo method: the independent perturbation approach. Ref. \cite{b7} also mentions the independent perturbation approach, but we do a more thorough analysis here.

In Section 2, we demonstrate theories of the basic resampling-based method and the independent perturbation method. Simultaneously, we show that the independent perturbation method reduces the variance of entries of the estimated FIM by from $O(1/N)$ to $O(1/(nN))$, where $n$ is the sample size of the data and $N$ is a measure of the Monte Carlo averaging. Section 3 illustrates the efficiency of the enhanced method through two numerical examples, respectively. The signal-plus-noise example verifies the theoretical ratio about variance reduction and compares the performance at each diagonal entries of the estimated FIM for the two methods. The mixture Gaussian example utilizes the relative error based on the spectral norm to show the obvious improvement of the independent perturbation method. Section 4 gives a conclusion and ideas about future work.

\section{Theoretical Analysis}
\subsection{Standard Method}

\quad We summarize aspects of the basic method in \cite{b5} below. Assume that $\pmb{Z}_\mathrm{pseudo}(i)$ is a set of $n$ independent random vectors generated by the Monte Carlo simulation that follow the distribution with parameters $\pmb{\theta}$, and $\hat{\pmb{H}}_{k|i}$ is the $k$th estimate of
$\pmb{H}(\pmb{\theta})$ at $\pmb{Z}_\mathrm{pseudo}(i)$. Additionally, suppose that $\pmb{\Delta}_{k|i}=\left[\Delta_{k1|i},\Delta_{k2|i},\ldots,\Delta_{kp|i}\right]^T$ is a random vector with a zero mean where the scalar elements are independent and identical symmetrically distributed random variables that are uniformly bounded and satisfy $\mathbb{E}\left(\left|1/\Delta_{kj|i}\right|\right)<\infty$, where $j=1,\ldots,p$, and $\pmb{\Delta}_{k|i}^{-1}\equiv[\Delta_{k1|i}^{-1},\Delta_{k2|i}^{-1},\ldots,\Delta_{kp|i}^{-1}]^T$. 

Here we generate an estimate of $\pmb{H}(\pmb{\theta})$ by simultaneous perturbation (SPSA) methods:
\begin{equation}
	\hat{\pmb{H}}_{k|i}=\frac{1}{2}\left[\frac{\delta \pmb{g}_{k|i}}{2c}(\pmb{\Delta}_{k|i}^{-1})^T+\left(\frac{\delta \pmb{g}_{k|i}}{2c}(\pmb{\Delta}_{k|i}^{-1})^T\right)^T\right],
\end{equation}
where $\delta \pmb{g}_{k|i}=\pmb{g}(\pmb{\theta}+c\pmb{\Delta}_{k|i}|\pmb{Z}_\mathrm{pseudo}(i))-\pmb{g}(\pmb{\theta}-c\pmb{\Delta}_{k|i}|\pmb{Z}_\mathrm{pseudo}(i))$, and $\pmb{g}(\cdot)$ is the (estimated if there is not enough information) gradient function of $L$, and $c>0$ is a small constant.  

The Monte Carlo-based estimate of $\pmb{F}(\pmb{\theta})$ in \cite{b5}, denoted $\bar{\pmb{F}}_{M,N}(\pmb{\theta})$, is
\begin{equation}
	\bar{\pmb{F}}_{M,N}(\pmb{\theta})\equiv-\frac{1}{N}\sum_{i=1}^N\frac{1}{M}\sum_{k=1}^M\hat{\pmb{H}}_{k|i}.
\end{equation}
The first “inner” average calculates Hessian estimates at a given $\pmb{Z}_\mathrm{pseudo}(i)$ ($i = 1, 2, \ldots, N$) based on $M$ values of $\hat{\pmb{H}}_{k|i}$, and the second “outer” average sums these sample means of Hessian estimates across $N$ values of pseudo data. 

To get the most intuitive comparison, we set $M=N=1$ without loss of generality since (2) indicates that Hessian estimates do not depend on $N$. Additionally, while $\hat{\pmb{H}}_{k|i} (k=1,\ldots,M)$ are estimated based on the same pseudo data with fixed $i$, \cite{b10} shows $M=1$ is the optimal choice if the $n$ vectors entering each $\pmb{Z}_\mathrm{pseudo}(i)$ are mutually independent. Then the variance of the $jj$th entry of the estimate $\bar{\pmb{F}}_{M,N}(\pmb{\theta})$ is 
\begin{equation}
	\mathrm{var}\left\{\left[\bar{\pmb{F}}_{M,N}\right]_{jj}\right\}=\mathrm{var}\left(\hat{H}_{jj}\right),
\end{equation}
where $\hat{H}_{jj}$ denotes the $jj$th entry of $\hat{\pmb{H}}=\hat{\pmb{H}}_{1|1}$. We can use the calculation at $M=N=1$ to readily extend to arbitrary matrix. Let $O_{(\cdot)}(c^2)$ denote a random “big-$O$” term, where the subscript represents the relevant randomness. For example, $O_{\pmb{Z},\pmb{\Delta}_1}(c^2)$ denotes a random “big-$O$” term depends on $\pmb{Z}_\mathrm{pseudo}(1)$ and $\pmb{\Delta}_1$ such that $O_{\pmb{Z},\pmb{\Delta}_1}(c^2)/c^2$ is bounded almost surely as $c\rightarrow0$. Then, by \cite{b8}, the $jj$th entry of $\hat{\pmb{H}}$ is 
\begin{equation}
	\hat{{H}}_{jj}={H}_{jj}+\sum_{l\neq j}{H}_{jl}\frac{\Delta_{1l}}{\Delta_{1j}}+O_{\pmb{Z},\pmb{\Delta}_1}(c^2).
\end{equation}
Assume that 
\begin{equation}
	\mathbb{E}(\Delta_{1l}/\Delta_{1j})=0,\; \mathrm{var}(\Delta_{1l}/\Delta_{1j})=v;
\end{equation}
where $v$ is a constant, then, given the independence of the $\left\{\Delta_{1j}\right\}$, and the fact that $\mathbb{E}({H}_{jl})=-{F}_{jl}$ ($j,l=1,2,\ldots,p$), we have:
\begin{align*}
	\mathrm{var}(\hat{{H}}_{jj})_{\text{basic}}=&\mathbb{E}\big([{H}_{jj}+\sum_{l\neq j}{H}_{jl}\frac{\Delta_{1l}}{\Delta_{1j}}+O_{\pmb{Z},\pmb{\Delta}_1}(c^2)]^2\big)-\big(\mathbb{E}[{H}_{jj}+\sum_{l\neq j}{H}_{jl}\frac{\Delta_{1l}}{\Delta_{1j}}+O_{\pmb{Z},\pmb{\Delta}_1}(c^2)]\big)^2\\
	=&\mathbb{E}\big({H}_{jj}^2+2{H}_{jj}\sum_{l\neq j}{H}_{jl}\frac{\Delta_{1l}}{\Delta_{1j}}+[\sum_{l\neq j}{H}_{jl}\frac{\Delta_{1l}}{\Delta_{1j}}]^2+O_{\pmb{Z},\pmb{\Delta}_1}(c^2)\big)-\big(\mathbb{E}\big[{H}_{jj}\big]+O(c^2)\big)^2\\
	=&\mathbb{E}({H}_{jj}^2)+\sum_{l\neq j}\mathbb{E}\bigg[{H}_{jl}\frac{\Delta_{1l}}{\Delta_{1j}}\bigg]^2-\big(\mathbb{E}\big[{H}_{jj}\big]\big)^2+O(c^2)\\
	=&\mathbb{E}({H}_{jj}^2)+v\sum_{l\neq j}\mathbb{E}({H}_{jl}^2)-{F}_{jj}^2+O(c^2)\\
	=&\mathrm{var}({H}_{jj})+v\sum_{l\neq j}\left[\mathrm{var}({H}_{jl})+{F}_{jl}^2\right]+O(c^2)\\
	=&\mathrm{var}({H}_{jj})+v\sum_{l\neq j}\mathrm{var}({H}_{jl})+v\sum_{l\neq j}{F}_{jl}^2+O(c^2).
\end{align*} 

\subsection{Implementation with Independent Perturbations}

\quad The estimation of $\pmb{F}(\pmb{\theta})$ can be enhanced if the $n$ vectors entering each $\pmb{Z}_\mathrm{pseudo}(i)$ are mutually independent. Note that the $n$ vectors do not have to be identically distributed --- just independent. By the independence, we have
$$\pmb{F}(\pmb{\theta})=\sum_{t=1}^n\pmb{F}^{(t)},$$
where $\pmb{F}^{(t)}$ denotes the FIM for each $t$.

In particular, the variance of the entries of each Hessian estimate $\hat{\pmb{H}}_{k|i}$ in the standard method can be reduced by decomposing
$\hat{\pmb{H}}_{k|i}$ into a sum of $n$ independent estimates $\hat{\pmb{H}}_{k|i}^{(t)}$, each corresponding to one vector in the pseudo data. To distinguish the way that $\pmb{\Delta}_{k|i}$ is generated in the basic method and the independent perturbation method, $\pmb{\Delta}_I$ is used here to represent the perturbation term $\pmb{\Delta}_{1|1}$ in the independent perturbation method, where $\pmb{\Delta}_I=\sum_{t=1}^n\pmb{\Delta}_I^{(t)}$. Then a separate perturbation vector $\pmb{\Delta}_I^{(t)}$ can be applied to each independent estimate, which results in variance reduction in the estimate $\bar{\pmb{F}}_{M,N}$. Hence, based on assumption (5), similar to the above calculation process, we have:
\begin{align*}
	&\mathrm{var}(\hat{{H}}_{jj})_{\text{indep}}=\mathrm{var}(\sum_{t=1}^n\hat{{H}}_{jj}^{(t)})=\sum_{t=1}^n\mathrm{var}(\hat{{H}}_{jj}^{(t)})\\
	=&\sum_{t=1}^n\{\mathbb{E}([{H}_{jj}^{(t)}+\sum_{l\neq j}{H}_{jl}^{(t)}\frac{\Delta_{Il}^{(t)}}{\Delta_{Ij}^{(t)}}+O_{\pmb{Z},\pmb{\Delta}_I}(c^2)]^2)-(\mathbb{E}[{H}_{jj}^{(t)}+\sum_{l\neq j}{H}_{jl}^{(t)}\frac{\Delta_{Il}^{(t)}}{\Delta_{Ij}^{(t)}}+O_{\pmb{Z},\pmb{\Delta}_I}(c^2)])^2\}\\
	=&\sum_{t=1}^n\{\mathbb{E}[({H}_{jj}^{(t)})^2]+v\sum_{l\neq j}\mathbb{E}[({H}_{jl}^{(t)})^2]-({F}_{jj}^{(t)})^2+O(c^2)\}\\
	=&\sum_{t=1}^n\{\mathrm{var}\left({H}_{jj}^{(t)}\right)+v\sum_{l\neq j}[\mathrm{var}({H}_{jl}^{(t)})+({F}_{jl}^{(t)})^2]+O(c^2)\}\\
	=&\sum_{t=1}^n\{\mathrm{var}({H}_{jj}^{(t)})+v\sum_{l\neq j}\mathrm{var}({H}_{jl}^{(t)})\}+v\sum_{t=1}^n[\sum_{l\neq j}({F}_{jl}^{(t)})^2]+O(c^2).
\end{align*}
The variance of the $j$th diagonal entry of $\hat{\pmb{H}}$ generated from the standard method can be written as: 
\begin{align*}
	\mathrm{var}(\hat{{H}}_{jj})_{\text{basic}}=&\mathrm{var}(\sum_{t=1}^n{H}_{jj}^{(t)})+v\sum_{l\neq j}\mathrm{var}(\sum_{t=1}^n{H}_{jl}^{(t)})+v\sum_{l\neq j}[\sum_{t=1}^n{F}_{jl}^{(t)}]^2+O(c^2)\\
	=&\sum_{t=1}^n\{\mathrm{var}\left({H}_{jj}^{(t)}\right)+v\sum_{l\neq j}\mathrm{var}({H}_{jl}^{(t)})\}+v\sum_{l\neq j}[\sum_{t=1}^n{F}_{jl}^{(t)}]^2+O(c^2).
\end{align*}
Therefore, the difference of the variance of the diagonal entries in the estimate of $\pmb{F}(\pmb{\theta})$ between the standard method and the independent perturbation method is
$$\mathrm{var}(\hat{{H}}_{jj})_{\text{basic}}-\mathrm{var}(\hat{{H}}_{jj})_{\text{indep}}=2v\sum_{l\neq j}\sum_{t_1<t_2}{F}_{jl}^{(t_1)}{F}_{jl}^{(t_2)}+O(c^2).$$ 

\subsection{Theoretical Comparison via Variance Reduction}

\quad Denoting $A=\sum_{t=1}^n\{\mathrm{var}({H}_{jj}^{(t)})+v\sum_{l\neq j}\mathrm{var}({H}_{jl}^{(t)})\}$, the result above shows that the variance difference between these two method is independent of $A$. That is, $\mathrm{var}(\hat{{H}}_{jj})_{\text{indep}}$ and $\mathrm{var}(\hat{{H}}_{jj})_{\text{basic}}$ contain the same components that are made up of the variance of entries of the true Hessian matrix. Hence, we should mainly consider the difference between $v\sum_{t=1}^n[\sum_{l\neq j}({F}_{jl}^{(t)})^2]$ and $v\sum_{l\neq j}[\sum_{t=1}^n{F}_{jl}^{(t)}]^2$. However, the result above cannot directly reflect the degree of variance reduction. Let us consider the ratio $\mathrm{var}(\hat{{H}}_{jj})_{\text{indep}}$ to $ \mathrm{var}(\hat{{H}}_{jj})_{\text{basic}}$, which can be written as:
\begin{equation}
	\frac{\text{var}(\hat{{H}}_{jj})_{\text{indep}}}{\text{var}(\hat{{H}}_{jj})_{\text{basic}}}=\frac{A+v\sum_{t=1}^n\left[\sum_{l\neq j}({F}_{jl}^{(t)})^2\right]+O(c^2)}{A+v\sum_{l\neq j}[\sum_{t=1}^n{F}_{jl}^{(t)}]^2+O(c^2)}.
\end{equation}
Usually $c$ is small (for example, 0.0001), so let us focus on the second part: off-diagonal entries of the true Fisher matrix. With a fixed $j$, let us for convenience denote entries ${F}_{jl}^{(t)}=f_l^{(t)}$, where $l\neq j$. Under some assumptions, the ratio  of the second part  in (6), $v\sum_{t=1}^n[\sum_{l\neq j}({F}_{jl}^{(t)})^2]/v\sum_{l\neq j}[\sum_{t=1}^n{F}_{jl}^{(t)}]^2$, is $O(1/n)$. That is, assume that:
\begin{enumerate}
	\item There are at least 2 non-zero off-diagonal entries in each true Fisher matrix $\pmb{F}^{(t)}$. 
	\item For every $j=1,2,\ldots,p$, the non-zero sequence $\{f_l^{(t)}\}_{t=1}^n$ is limited within $[f_{l}^{(\text{min})},f_{l}^{(\text{max})}]$ while fixing $l$, where $f_{l}^{(\text{min})}$ is the lower bound of non-zero entries in $\{|f_l^{(t)}|\}_{t=1}^n$, and $f_{l}^{(\text{max})}$ is the upper bound of $\{|f_l^{(t)}|\}_{t=1}^n$. For convenience, denote $r_{1_{j}}=\sum_{l\neq j}^p(f_{l}^{(\text{max})}/{f_{l}^{(\text{min})}})^2$. 
	\item Denote $f_l^{(+)}$ as the sum of all positive elements in the sequence, and $f_l^{(-)}$ as the sum of all negative elements in the sequence, the ratio is $r_{2_{j}}=(f_l^{(+)}+|f_l^{(-)}|)^2/{(f_l^{(+)}-|f_l^{(-)}|)^2}$ such that $r_{1_{j}}r_{2_{j}}=O(1)$ as $n$ goes to infinity. 
\end{enumerate}

Based on the above conditions, we have:

\textbf{Proposition 1:} Suppose that conditions (1), (2), and (3) hold for all $j$. Then the independent perturbations above reduce the variance of the diagonal entries in the estimate of $\pmb{F}(\pmb{\theta})$ by $O(1/n)+O(c^2)$.

The details about the proof have been shown in \cite{b10}, so we get:
\begin{align*}
	\frac{\text{var}(\hat{{H}}_{jj})_{\text{indep}}}{\text{var}(\hat{{H}}_{jj})_{\text{basic}}}
	\leq&\frac{A+\frac{\text{C}}{n}B_{\text{basic}}+O(c^2)}{A+B_{\text{basic}}+O(c^2)},
\end{align*}
where $C=r_{1_{j}}r_{2_{j}}$ is a constant, $A=\sum_{t=1}^n\{\mathrm{var}({H}_{jj}^{(t)})+v\sum_{l\neq j}\mathrm{var}({H}_{jl}^{(t)})\}=O(n)$, and $B_{\text{basic}}=v\sum_{l\neq j}^p[\sum_{t=1}^nf_{l}^{(t)}]^2\leq n^2v\sum_{l\neq j}^p[f_{l}^{(\text{max})}]^2=O(n^2)$. Therefore,  
\begin{equation}
	\frac{\text{var}\left(\hat{{H}}_{jj}\right)_{\text{indep}}}{\text{var}\left(\hat{{H}}_{jj}\right)_{\text{basic}}}=O\bigg(\frac{1}{n}\bigg)+O(c^2),
\end{equation}
where $O(c^2)$ is often so small that it can be ignored (see Example 1 in Section 3).

In addition, similar to Section 4.2 in \cite{b5}, \cite{b10} shows that in the independent case, $M>1$ is considered only when it is costly to generate new pseudodata vectors. This is despite the fact that we mentioned earlier that the average of $M$ estimated Hessian matrices, which is based on the same pseudodata vector $\pmb{Z}_{\text{pseudo}}$, is usually used to estimate a Fisher information matrix.

That is, in general, if it is allowed to compute $C=MN$ Hessian matrix estimates, we can maximize the accuracy of $\bar{\pmb{F}}_{M,N}$ when each estimate $\hat{\pmb{H}}_{k|i}$ is generated based on a new $\pmb{Z}_{\text{pseudo}}$ (i.e., $M=1$).

\textbf{Proposition 2:} Assume that elements $\pmb{\Delta}_{k|i}^{(t)}$ and $\pmb{Z}^{(t)}_{\text{pseudo}}(i)$ are mutually independent, where $k=1,\ldots,M$, $i=1,\ldots,N$, and $t=1,\ldots,n$. For a fixed $C=MN$, the variance of each entry in $\bar{\pmb{F}}_{M,N}(\pmb{\theta})$ is minimized when $M=1$.

\section{Numerical Examples}

\quad Here we introduce two examples, which are also used in \cite{b11}. The first one is under the signal-plus noise setting. We calculate the variance ratio to verify the theoretical result and analyze the distinction among the variance ratio for different diagonal entries. The second one compares the performance for the two methods when the pseudo data are mixture Gaussian distributed. At the same time, we show the improvement of the enhanced method from time analysis in both examples.

Both of the mixture distribution and the signal-plus-noise settings arise regularly in the literature. Mixture problems are thoroughly reviewed in \cite{b12} and \cite{b13}, with applications in information theory in \cite{b14} and bivariate quantile estimation in \cite{b15}. The signal-plus-noise problem with non-identical noise distributions arises in practical problems where measurements are collected with varying quality of information across the sample. Some practical implications are discussed in \cite{b16} relative to the initial conditions in a Kalman filter model and \cite{b17} in the context of outlier analysis.

\subsection{Multivariate Normal with Signal-Plus-Noise}

\quad Assume that the $\pmb{z}_i$ are independently and normally distributed with mean $\pmb{\mu}$ and covariance $\pmb{\Sigma}+\pmb{P}_i$ for all $i$, where $\pmb{\mu}$ and $\pmb{\Sigma}$ are unknown parameters to be estimated. It can be interpreted that $\pmb{z}_i$ is the observed values of the $N(\pmb{\mu},\pmb{\Sigma})$ distributed signal with independent $N(\pmb{0},\pmb{P}_i)$ distributed noise. This setting has been widely discussed; for example, \cite{b16} considers estimating the initial mean vector and covariance matrix of a state-space model and \cite{b18} demonstrates parameter estimation of random-coefficient linear models.

Let us consider a 3-dimensional case: $\text{dim}(\pmb{z}_i)=3$, $n=30$, $\pmb{\mu}=\pmb{0}$, $\pmb{\Sigma}$ is a matrix with all 0.5's except 2's on the diagonal (which is a little different from that in \cite{b10}), and $\pmb{P}_i=\sqrt{i}\pmb{U}^T\pmb{U}$, where $\pmb{U}$ is a $3\times3$ matrix with entries generated from uniform (0,1) distribution. $\pmb{\theta}$ is the collection of unique elements in $\pmb{\mu}$ and $\pmb{\Sigma}$, so the FIM is $p\times p$ where $p=9$. Reference \cite{b16} gives the gradient function of log-likelihood and the analytical form of the FIM in this setting. 

In particular, assume that components of the perturbation vector $\pmb{\Delta}_{k|i}$ follow Bernoulli $\pm1$ distribution. In both methods, we estimate the Hessian matrix by using the gradient of the log-likelihood function under $M=N=1$. The ratio values for estimated variance of the diagonal entries of the estimated FIM from the standard method to that from the independent perturbation method based on 2 million replicates are shown in Table 1. Furthermore, we change $n$ from 30 to 200 to observe the performance.

\begin{table}[htbp]
	\caption{Variance of $\bar{\pmb{F}}_{M,N}(\pmb{\theta})_{jj}$}
	\begin{center}
		\begin{tabular}{|c|c|c|c|}
			\hline
			\textbf{Diagonal}& \textbf{Independent} & \textbf{Standard} & \textbf{Ratio}\\
			\hline
			$j=1$ &8.24  &60.12 & 0.14\\
			\hline
			$j=2$ &13.32 &56.58 &0.23\\
			\hline
			$j=3$ & 4.88 & 24.48& 0.20\\
			\hline
			$j=4$ & 5.00 & 	11.18 &0.45\\
			\hline
			$j=5$ &23.49 & 39.13 & 0.60\\
			\hline
			$j=6$ &14.85 &26.69 & 0.56\\
			\hline
			$j=7$ &8.69 & 13.84 & 0.63\\
			\hline
			$j=8$ &12.91 & 19.70 & 0.65\\
			\hline
			$j=9$ &3.04 &4.98 & 0.61\\
			\hline
		\end{tabular}
		\label{tab1}
	\end{center}
\end{table} 

Table 1 indicates that there is an obvious reduction of the variance of entries in the estimated FIM over the enhanced method compared with that of the standard method for every $j$, which matches our theoretical conclusion in Section 2.

\begin{table}[htbp]
	\caption{Time cost/$seconds$}
	\begin{center}
		\begin{tabular}{|c|c|c|c|}
			\hline
			&\textbf{$n=30$}& \textbf{$n=100$}&\textbf{$n=200$}\\
			\hline
			Independent Method &8726&	28914&	57066			\\
			\hline
			Standard Method &5259&	16978&	33541\\
			\hline
			Ratio & 0.6027&	0.5872	&0.5878\\
			\hline
		\end{tabular}
		\label{tab1}
	\end{center}
\end{table} 

On the other hand, researchers usually need to trade off accuracy and time cost in experiments. Results with high accuracy generally take more time to obtain, but the payoff may not be worthwhile if the computational cost is too high. Therefore, under the same setting as above, Figures 1 and 2, and Table 2 show the performance of the standard method and the enhanced method as $n$ is equal to 30, 100, and 200 based on the variance ratio $\text{var}(\hat{H}_{jj})_{\text{indep}}/\text{var}(\hat{H}_{jj})_{\text{basic}}$ and time cost. To better compare the variance ratio with $O(1/n)$ as $n$ grows, we add the reference lines $3.5/n$ and $13/n$ in Figures 1 and 2, separately. Figure 1 and 2 reveal that the change of variance ratio as $n$ grows is similar to the trend of $O(1/n)$. That is, the curves follow the overall slope given by the reference line. But the time cost ratio in Table 2 is stable over the range of $n$ considered. 

\begin{figure}[H]
	\centering
	\includegraphics[width=9.5cm,height=7.3cm]{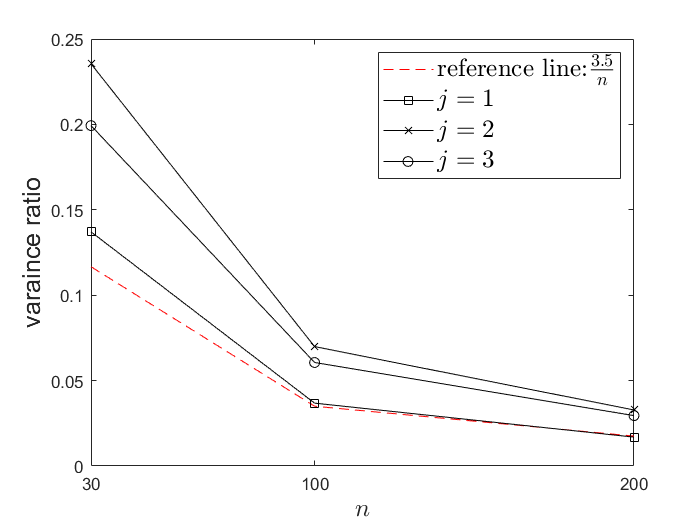}
	\caption{Variance ratio for $j=1,2,3$}
\end{figure}

\begin{figure}[H]
	\centering
	\includegraphics[width=9.5cm,height=7.3cm]{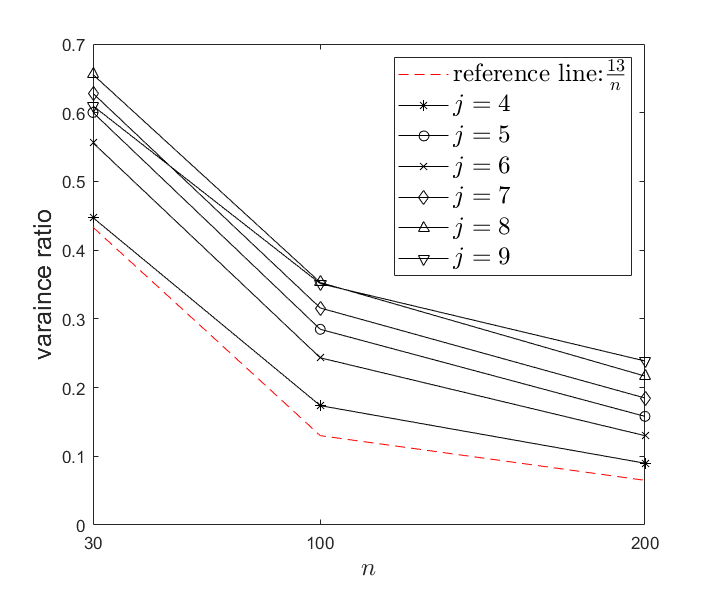}
	\caption{Variance ratio for $j=4,\ldots,9$}
\end{figure}

 In addition, under specific regularity conditions \cite{b4}, the $jj$th entry of $\pmb{F}(\pmb{\theta})$ is the expectation of $\partial^2 L/\partial {\theta}_j^2$, where ${\theta}_j(j=1,\ldots,9)$ corresponds to three elements in $\pmb{\mu}$ and six unique entries in $\pmb{\Sigma}$ respectively. We can see that the estimates of diagonal entries of $\pmb{F}(\pmb{\theta})$ corresponding to $\pmb{\mu}$ part are more stable than those corresponding to the $\pmb{\Sigma}$ part. Furthermore, when $n=30$, the variance ratio values in Figure 1 are significantly lower than 0.6027, which is the time cost ratio in Table 2. Part of the variance ratio values in Figure 2, however, are higher than 0.6027. 

Note that compared with \cite{b10}, we only change the diagonal values of the true covariance matrix $\pmb{\Sigma}$. But in \cite{b10}, all variance ratio values are much lower than 0.9586, which is the time cost ratio as $n=30$. Here, the performance of variance ratios that corresponds to the $\pmb{\mu}$ part is much better than that corresponding to the $\pmb{\Sigma}$ part. That is, the variance ratio values corresponding to the $\pmb{\mu}$ part are lower and closer to $1/n$. This might be because the true variance value for each dimension becomes large, which tend to expand the range of generated values and affects the estimation for the covariance part. Then the fluctuation of corresponding diagonal entries (for the covariance part) in the estimated FIM are relatively obvious compared with that for the mean part.

Although Table 2 indicates that it may not be worth the time (compare the time cost ratio with part of the variance ratio values) to improve accuracy when $n=30$, the results for $n=100, 200$ from the independent method achieve a significant reduction on variance with acceptable and reasonable time cost, which means that our enhanced method is applicable in practice.

\subsection{Mixture Gaussian Distribution}

\quad Denote the generated vector $\pmb{Z}=[z_1,z_2,\ldots,z_n]^T$, and assume that $z_i \;(i=1,\ldots,n)$ are independently and identically distributed with probability density function:
$$f(z,\pmb{\theta})=\lambda\exp(-(z-\mu_1)^2/(2\sigma_1^2))/\sqrt{2\pi\sigma_1^2}+(1-\lambda)\exp(-(z-\mu_2)^2/(2\sigma_2^2))/\sqrt{2\pi\sigma_2^2},$$
where $\pmb{\theta}=[\lambda,\mu_1,\sigma_1^2,\mu_2,\sigma_2^2]^T$. There are $5(5+1)/2=15$ unique entries in the FIM. Furthermore, we use the average of the negative Hessian matrix based on the sample vector over $10^6$ independent replicates to approximate (since \cite{b19} showed that the Hessian matrix is attainable) the true FIM, which is not computable in this example.

Here consider the same setting as that in \cite{b11}, where $\pmb{\theta}=[0.2,0,4,1,9]^T$, $M=2$, $N=40000$, $c=0.0001$, and elements of the perturbation $\pmb{\Delta}_{k|i}$ follow Bernoulli $\pm1$ distribution for all $k\;(k=1,\ldots,N)$ and $i\;(i=1,\ldots,M)$. We estimate the true FIM $\pmb{F}(\pmb{\theta})$ given the gradient of the log-likelihood function, and measure the performance of methods by computing the sample mean of $\|\pmb{F}_{\text{est}}(\pmb{\theta})-\pmb{F}(\pmb{\theta})\|/\|\pmb{F}(\pmb{\theta})\|$ based on 50 independent replicates, where $\|\bullet\|$ represents the spectral norm. 
Table 3 gives the typical estimated FIM based on the standard method and the independent method, respectively. By typical, we mean it is the estimated FIM corresponding to the 25th value of $\|\pmb{F}_{\text{est}}(\pmb{\theta})-\pmb{F}(\pmb{\theta})\|/\|\pmb{F}(\pmb{\theta})\|$ in descending order given 50 independent replicates. Note that the true FIM is approximated by sample averages over $10^6$ independent replicates. Table 4 summarizes the performances (accuracy and time cost) between the standard method and the independent approach.
 
\begin{table}[htbp]
	\caption{Simulation Results for Example 3.2}
	\begin{center}
		\begin{tabular}{|c|c|}
			\hline
			True FIM&$\left[
			\begin{array}{ccccc}796.8750&	0&	0&	0&	0\\
				0&	7.5000&	1.5000&	0&	0\\
				0&	1.5000&	3.1876&	0&	0\\
				0&	0&	0&	3.3333&	-0.0741\\
				0&	0&	0&	-0.0741&	0.1523\end{array}
			\right]
			$\\
			\hline
			Typical $\bar{\pmb{F}}(\pmb{\theta})_{\text{basic}}$&$\left[
			\begin{array}{ccccc}
				796.8752	&1.5414&	1.8820	&0.3499&	-0.7470\\
				1.5414&	7.5004&	1.4958&	0.01550&	-0.0075\\
				1.8820&	1.4958&	3.1868&	0.0157&	-0.02251\\
				0.3499	&0.0155	&0.0157&	3.3332&	-0.0863\\
				-0.7470&	-0.0075&	-0.02251&	-0.0863&	0.1521
			\end{array}
			\right]
			$\\
			\hline
			Typical $\bar{\pmb{F}}(\pmb{\theta})_{\text{indep}}$&$\left[
			\begin{array}{ccccc}
				796.8752&0.2536&	0.0278&	0.4273&	-0.0930\\
				0.2536	&7.4949&	1.4922&-0.0043&	-0.0009\\
				0.0278&1.4922&	3.1768	&-0.0019&	-0.0026\\
				0.4273	&-0.0043&	-0.0019&3.3323	&-0.0754\\
				-0.0930&	-0.0009&	-0.0026&	-0.0754&	0.1509
			\end{array}
			\right]$
			\\
			\hline
		\end{tabular}
		\label{tab1}
	\end{center}
\end{table}

\begin{table}[htbp]
	\caption{Comparison between Two Methods}
	\begin{center}
		\begin{tabular}{|c|c|c|c|}
			\hline
			&Independent &Standard&Ratio\\
			\hline
			Accuracy ($\|\pmb{F}_{\text{est}}(\pmb{\theta})-\pmb{F}(\pmb{\theta})\|/\|\pmb{F}(\pmb{\theta})\|$) &0.00063&	0.00330&	0.19\\
			\hline
			Time Cost/$seconds$&1760&	1212&	1.45\\
			\hline
		\end{tabular}
		\label{tab1}
	\end{center}
\end{table} 
 Table 4 shows the obvious advantage of the enhanced method with independent perturbations while taking time cost into account. It is worth noting that our accuracy of the basic approach is higher than that shown in \cite{b11} because for each time, i.e., fixing $k$ and $i$, \cite{b11} only generates one sample following the mixture Gaussian distribution, but we generate $n$ samples, which might result in a more accurate estimation.

\section{Conclusion and Future Work}

\quad It is often hard to estimate the FIM in real-world models because of high dimension or nonlinear functions. This paper presents an enhanced Monte Carlo method with independent perturbations to estimate the Fisher information matrix in problems where it is not analytically available. The theoretical analysis and numerical examples show that this approach improves the estimation performance by reducing variance of elements compared to the basic Monte Carlo approach.

Moreover, the above results indicate that the performance might change even if we only change certain parameters in the model; for example, consider the time cost ratios in Example 1 and \cite{b10}. However, as $n$ increases, the advantage of the enhanced method is apparent with no doubt. 

In future work, in addition to considering the performance of estimated FIM in other practical problems (for example, Bayesian statistics \cite{b20} and the EM algorithm \cite{b8}), it is important to account for the relative benefit and cost of reducing variance, especially when $n$ is very large, where we cannot ignore the cost of generating the additional perturbation vectors in the independent perturbation method.
\begin{singlespace}
	
\end{singlespace}
\end{document}